\documentclass{aa}  

\usepackage{graphicx}
\usepackage{xcolor}
\usepackage{gensymb}
\usepackage{ulem}
\usepackage{multirow}
\usepackage{comment}
\newcommand{\new}[1]{\textcolor{black}{ #1}}
\usepackage{txfonts}
\begin{document}

   \title{Discovery of two new millisecond pulsars towards the Galactic bulge}

   \author{J. Berteaud\inst{1,2}
          \and
          F. Calore\inst{3}
          \and
          M. Clavel\inst{4}
          \and
          S. Dai\inst{5,6}
          \and
          J. S. Deneva\inst{7}
          \and
          S. Hyman\inst{8}
          \and
          F.K. Schinzel\inst{9}\thanks{An Adjunct Professor at the University of New Mexico.}
          \and
          A. Ridolfi\inst{10}
          \and
          S. M. Ransom\inst{11}
          \and
          F. Abbate\inst{12,13}
          \and
          C. J. Clark\inst{14,15}
          \and
          M. Kramer\inst{13,16}
          \and
          T. Thongmeearkom\inst{16,17}
          \and
          B. W. Stappers\inst{16}
          \and
          E. D. Barr\inst{13}
          \and
          R. P. Breton \inst{16}
          }

   \institute{
   NASA Goddard Space Flight Center, Code 662, Greenbelt, MD 20771, USA\\
    \email{joanna.n.berteaud@nasa.gov}
    \and
    University of Maryland, Department of Astronomy, College Park, MD 20742, USA
    \and
    LAPTh, CNRS, USMB, F-74940 Annecy, France
    \and
    Université Grenoble Alpes, CNRS, IPAG, F-38000 Grenoble, France
    \and
    Australia Telescope National Facility, CSIRO, Space and Astronomy, PO Box 76, Epping, NSW 1710, Australia
    \and
    Western Sydney University, Locked Bag 1797, Penrith South DC, NSW 2751, Australia
    \and
    George Mason University, resident at the Naval Research Laboratory, Washington, DC 20375, USA
    \and
    Department of Engineering and Physics, Sweet Briar College, Sweet Briar, VA 24595, USA
    \and
    National Radio Astronomy Observatory, P.O. Box O, Socorro, NM 87801, USA
    \and
    Fakult\"at f\"ur Physik, Universit\"at Bielefeld, Postfach 100131, D-33501 Bielefeld, Germany
    \and
    National Radio Astronomy Observatory, 520 Edgemont Rd., Charlottesville, VA 22903, USA
    \and INAF – Osservatorio Astronomico di Cagliari, Via della Scienza 5, 09047 Selargius (CA), Italy
    \and Max-Planck-Institut f\"ur Radioastronomie, Auf dem Hu\"ugel 69, D-53121 Bonn, Germany
    \and 
    Max Planck Institute for Gravitational Physics (Albert Einstein Institute), D-30167 Hannover, Germany
    \and Leibniz Universit\"at Hannover, D-30167 Hannover, Germany
    \and Jodrell Bank Centre for Astrophysics, Department of Physics and Astronomy, The University of Manchester, Manchester M13 9PL, UK
    \and
    National Astronomical Research Institute of Thailand, Don Kaeo, Mae Rim, Chiang Mai 50180, Thailand
        }

   \date{}
 
  \abstract
    {}
   {The mysterious Galactic Center Excess of $\gamma$ rays could be explained by a large population of millisecond pulsars hiding in the Galactic bulge, too faint to be detected as individual high-energy point sources by the \textit{Fermi} Large Area Telescope, as well as too fast and too dispersed to be detected in shallow radio pulsation surveys.}
   {Motivated by an innovative candidate selection method, we aim at detecting millisecond pulsars associated with the Galactic Center Excess by carrying deep radio pulsation searches towards promising candidates detected in the inner Galaxy, in X rays by \textit{Chandra}, and in radio or $\gamma$ rays by the Very Large Array or \textit{Fermi}.}
   {We conducted deep radio observation and follow-up campaigns with MeerKAT, the Murriyang and the Green Bank telescopes towards 9 X-ray candidate sources.} 
   {We here report the detection of two new millisecond pulsars\new{, including a black widow candidate, towards the Galactic bulge: PSRs J1740--2805 and J1740--28}. These discoveries double the number of MSPs discovered within the innermost $2\degree$ from the Galactic center.}
   {}

   \keywords{pulsars: individual}

   \maketitle

\section{Introduction}
\label{sec:intro}

    The Galactic Center Excess (GCE) discovered by the $\gamma$-ray Large Area Telescope (LAT) aboard the \textit{Fermi} satellite has been puzzling astrophysicists for more than fifteen years \citep[see][for a review]{2020ARNPS..70..455M}. Since its discovery, two main explanations have been put forward: dark matter annihilation, and unresolved millisecond pulsars (MSPs), i.e., faint pulsars with rotation periods shorter than 30 ms. Dark matter is indisputably present in the Milky Way, and possibly strongly peaked at the Galactic center \citep{2021PDU....3200826B, 2024MNRAS.528..693O}, but whether or not it self-annihilates into $\gamma$-ray photons in the \textit{Fermi}-LAT energy band and produces a spectrum comparable to the GCE spectrum depends on the nature of dark matter itself and on its properties. Hence, identifying the origin of the GCE could have a strong impact on our understanding of dark matter. Alternatively, the GCE could be caused by yet undetected MSPs. Its bulge-like morphology \citep{2018NatAs...2..819B, 2018NatAs...2..387M, 2022ApJ...929..136P, 2024MNRAS.530.4395S} and its photon-count statistics \citep{2021PhRvL.127p1102C,2024PhRvD.109l3042M,2021PhRvD.104l3022L,2022PhRvD.105f3017M}, among other pieces of evidence, strongly hint at a stellar origin of the GCE, at least partial, therefore favoring the MSP hypothesis. The existence of a large population of MSPs in the Galactic bulge has sometimes been denied in the context of recycled MSP formation \citep{2017JCAP...05..056H, 2022MNRAS.512.4239B}, but can be explained by invoking other formation mechanisms, as for example accretion-induced collapse of O-Ne white dwarfs in binary systems \citep{2022NatAs...6..703G} and disrupted globular clusters \citep{2022ApJ...940..162Y}. Unlike MSPs in globular clusters and in the Galactic disk, evidence for MSPs in the Galactic bulge is not yet supported by individual radio detections. \new{According to some recent models, the surface density of disk pulsars drastically decreases towards the GC; however, other models predict a residual population within the inner 500 pc \citep{2024ApJ...963L..39X}, while others don't \citep{2018MNRAS.481.3966B}. But in all models, t}he bulge and the disk overlap in the inner Galaxy; associating MSPs with either of these components based on their position alone is therefore challenging. Moreover, if the position of MSPs on the celestial sphere is often known with very good accuracy, their distance, typically computed from their dispersion measure (DM), is more uncertain, and strongly dependent on models of the free-electron density of the Galaxy. Existing radio pulsation surveys\footnote{See \url{https://www.jb.man.ac.uk/pulsar/surveys.html} for a list of major pulsar surveys conducted since 1967.} in the GHz band of the inner Galaxy \citep[see][and references therin]{2016MNRAS.459.1104L} are highly relevant to discover the least dispersed, i.e., closest disk or bulge MSPs, but are too shallow to unveil distant MSPs, undeniably associated with the bulge and the GCE \citep{2016ApJ...827..143C}.
    Finding MSPs further in, towards the Galactic center region would not only be useful for understanding stellar evolution and the GCE, but also for probing the spacetime and around Sagittarius A$^*$ \citep[e.g.,][]{2012ApJ...747....1L} and the density of free electrons in the inner Galaxy \citep[e.g.,][]{2017ApJ...835...29Y}. More generally, MSPs are highly relevant for the study of the gravitational wave background from supermassive black holes mergers, \cite[e.g.,][]{2023ApJ...951L...8A,2023A&A...678A..50E} and the equation of state of cold, dense matter \citep[e.g.,][]{2022NatRP...4..237Y}.

        \begin{table*}[t]
            \caption{Summary of our observations.}
            \begin{center}
                \begin{tabular}{c|c|c|c|c|c}
                    \hline
                    \hline
                    2CXO name & Galactic long. & Galactic lat. & P1152 observations & 22B-112 observations & MeerKAT observations\\
                     & (deg) & (deg) & Duration, date & Duration, date & Duration, date\\
                    \hline
                    J173946.6--282913 & 359.70782 & 1.32561 & 3 h, 22--03--31 & 1 h, 22-11-11 & --- \\ 
                    J174053.7--275708 & 0.29178 & 1.40042 & 8 h, 22--04--17 & 1 h, 22--11--11 & --- \\ 
                    J174007.6--280708 & 0.06043 & 1.45587 & --- & 3 h, 23--01--02 & --- \\
                    J174011.5--283221 & 359.71181 & 1.22051 & 6 h, 22--05--25 & 2.5 h, 23--07--11 & --- \\
                    J174017.3--282843 & 359.77436 & 1.23492 & --- & 3.75 h, 22-12-24 & --- \\
                    J174306.8-293344 & 359.18046 & 0.13807 & --- & 5.25 h, 22--08--12 & --- \\
                    J173545.5--302859 & 357.59 & 1.02 & --- & --- & 2 h, 23--09--21 \\
                    J174235.4--282829 & 359.35 & --1.66 & --- & --- & 1.5 h, 23--05--20\\
                    J175039.6--302056 & 0.045 & 0.81 & --- & --- & 1.5 h, 23--05--20\\
                    \hline
                \end{tabular}
            \tablefoot{From left to right, the columns give the X-ray target name, its longitude and latitude as well as details on the initial Murriyang (P1152), GBT (22B-112) and MeerKAT observations. The date format is YY--MM-DD.}
            \end{center}

            \label{tab:obs}
    \end{table*}
    
    The MSP emission spectrum is truly MW, making it possible to search for their emission across frequencies. The launch of the \textit{Fermi}-LAT enabled a large number of $\gamma$-ray pulsar discoveries over the last decade, and more than 300 are known to date\new{, to include more than 140 MSPs} \citep[see, e.g.,][and references therein]{2023ApJ...958..191S}. Recently, new radio facilities like MeerKAT and FAST, the Five-hundred-meter Aperture Spherical Telescope, have greatly contributed to increase the detection rate by targeting, among others, unidentified \textit{Fermi} sources \citep[e.g.,][]{2021SCPMA..6429562W, 2023MNRAS.519.5590C}. In our multi-wavelength quest for MSPs, we previously showed that $\sim10^2$ GCE MSPs could have been detected in X-ray imaging observations of the Galactic center \citep{2021PhRvD.104d3007B}, assuming that bulge MSPs are the sole cause of the GCE. Then, we identified a large population of MSP candidates within the \textit{Chandra} Source Catalog \citep[CSC,][]{2010ApJS..189...37E} with multi-wavelength properties in agreement with expectations for bulge MSPs. Among these sources, 9 have either a radio or a $\gamma$-ray plausible counterpart, making them promising MSP candidates \citep{2024A&A...690A.330B}. We conducted deep follow-up radio observations of these 9 sources potentially associated with the GCE using MeerKAT, the Murriyang and the Green Bank telescopes and here report on the discovery of two new MSPs towards the Galactic bulge. We describe our observations and results in Sections \ref{sec:obs} and \ref{sec:obs_res}. Discussion and conclusions are presented in Sections \ref{sec:discussion} and \ref{sec:conclusion}.
   
\section{Observations and data reduction}
\label{sec:obs}

        \subsection{Radio-motivated targets}

        In \cite{2024A&A...690A.330B}, we identified 5 sources only detected at X-ray and radio wavelengths, by \textit{Chandra} and the Karl G. Jansky Very Large Array (VLA) of the National Radio Astronomy Observatory: 2CXO J173946.6--282913, 2CXO J174053.7--275708, 2CXO J174007.6--280708, 2CXO J174011.5--283221 and 2CXO J174017.3--282843. These sources are all within $4\degree$ of the Galactic center. We looked for and did not find any known bright infrared, optical or ultraviolet counterparts. MSPs are usually faint at these wavelengths, even more so in the Galactic bulge, i.e., if they are distant. These 5 sources are therefore suitable MSP candidates and promising targets for deep radio pulsation searches. We also observed an additional \textit{Chandra} source having a VLA counterpart: 2CXO J174306.8--293344. This source was rejected as an MSP candidate in \cite{2024A&A...690A.330B} because of a positive match with an infrared source located on the rim of the 95\% error circle of the X-ray source position. We considered this match as a false positive and included this source in our observing program. We provide more information on 2CXO J174306.8--293344 in Appendix \ref{app:J1743}.

        \subsubsection{Initial observations}

        We conducted initial observations of these 6 pulsar candidates with Murriyang, CSIRO's Parkes Radio Telescope and the Green Bank Telescope (GBT) of the Green Bank Observatory. We planned our observations accounting for the sources' flux density at L-band, measured from the VLA image.
        We observed three of the pulsar candidates with the Murriyang telescope and its Ultra-Wide-bandwidth Low-frequency \citep[UWL, 704--4032~MHz,][]{2020PASA...37...12H} receiver system during semester 2022APR (Project P1152). Observing time varied from 3 to 8 hours. Pulsar search-mode data were recorded with the MEDUSA backend, a 2-bit sampling, a sampling time of 64 $\mu$s and 2048 channels for each of the 26 subbands (128 MHz per subband).
        We observed all 6 pulsar candidates with the GBT at C (2CXO J174306.8--293344, 3.95--7.8 GHz) or S-band (5 others, 1.73--2.60 GHz) during semester 2022B (Project 22B-112). Observing time varied from 1 to 5 hours. Pulsar search-mode data were recorded using the VEGAS backend. For S-band, we used a bandwidth of 800~MHz with 2048 channels and 82~$\mu$s sampling time. For C-band we used four VEGAS boards, each with a bandwidth of 1500~MHz, 4096 channels, and 87~$\mu$s sampling time. Our initial observations are summarized in Table \ref{tab:obs}.

    \subsubsection{Pulsation search}

        The pulsar search-mode data were searched for periodic dispersed radio pulsations using \texttt{PRESTO}\footnote{\url{https://github.com/scottransom/presto}} \citep{2011ascl.soft07017R}. Our analysis includes the standard RFI excision, Fourier-domain acceleration search \new{(but no jerk search)} and optimization for signals with changing apparent spin periods caused by orbital motion \citep{2002AJ....124.1788R}. We accounted for DMs in the range 0 to 1000~cm$^{-3}$~pc.
        Each Murriyang observation was split into three 1024 MHz-wide bands, LOW ($\sim$0.9 GHz to $\sim$1.9 GHz), MID ($\sim$2 GHz to $\sim$3 GHz) and HIGH ($\sim$3 GHz to $\sim$4 GHz) and processed separately.
        GBT C-band data from the four VEGAS boards were  combined offline resulting in a combined data set of 4500~MHz bandwidth, split in 12288 channels.

        \begin{figure}[ht]
            \centering
            \includegraphics[width=0.4\textwidth, trim={0 2cm 0 3cm}, clip]{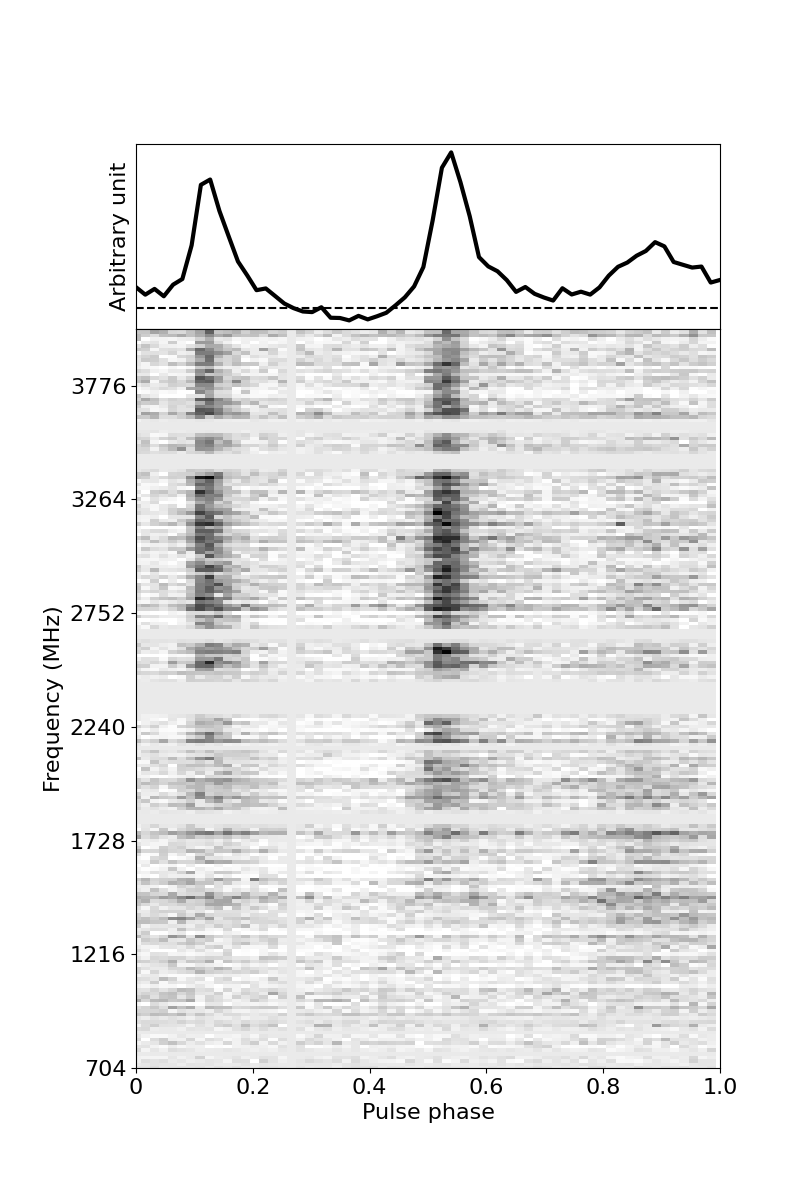}
            \includegraphics[width=0.4\textwidth, trim={0 2cm 0 3cm}, clip]{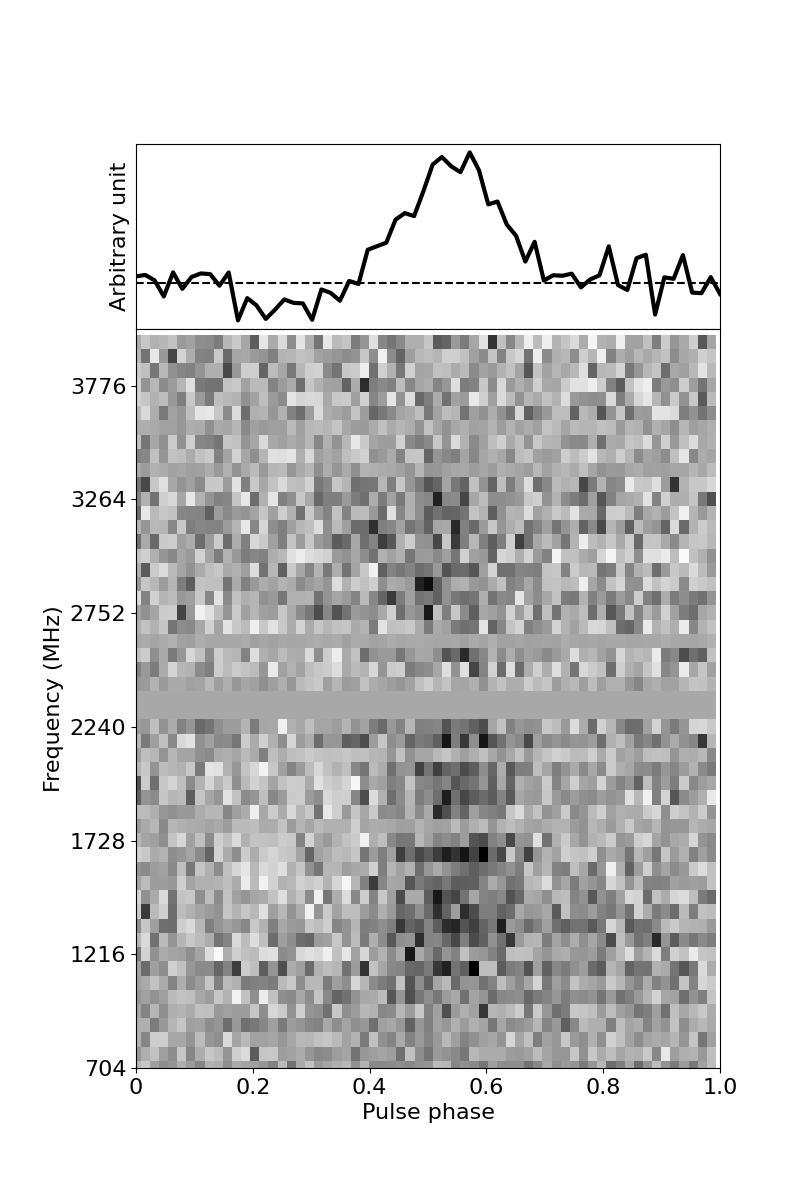}
            \caption{Integrated pulse profile and frequency-phase waterfall of PSRs~J1740--2805 (top) and J1740--28 (bottom). Data from all available Murriyang observing epochs were combined to enhance the signal-to-noise ratio.}
            \label{fig:pulse_profiles}
        \end{figure}

    \subsubsection{Timing and positions}

        We detected a pulsar in the direction of 2CXO J174007.6--280708 (see Sect. \ref{sec:obs_res}), one of the MSP candidates targeted with the GBT, and immediately sought and obtained Director's time at the VLA (Project 23B-312) to investigate the precise position of the newly discovered pulsar in imaging data. The VLA observation indicated that the pulsar and 2CXO J174007.6--280708 were likely not associated\new{, as no continuum radio source was detected at the position of the X-ray target}. More details on the VLA data analysis are provided in Appendix \ref{app:vla}. We also sought and obtained Parkes Director's time (Project PX117), with which we confirmed the detection of PSR J1740--28 but also identified and confirmed the detection of a second pulsar in the same beam: PSR J1740--2805. \new{The latter was then detected in the VLA observation mentioned above (see Appendix \ref{app:vla} for more details), and is also unassociated with our initial X-ray target.}
    
        A regular timing campaign of PSRs~J1740--28 and J1740--2805 was started during semester 2024APR, regularly observing their position with a single beam of the Murriyang telescope for 4 hours per epoch (Project P1317). Both pulsar having a similar DM (see Sect. \ref{sec:obs_res}), the UWL system was used in the coherently de-dispersed search mode with ${\rm DM}=317$\,pc\,cm$^{-3}$, the DM of the faintest of both pulsars, and 1\,MHz frequency resolution. Each observation was folded using the \texttt{DSPSR}\footnote{\url{https://dspsr.sourceforge.net/}}~\citep{2011PASA...28....1V} software package with a sub-integration length of 30\,s. Data affected by narrow-band and impulsive {RFI} were manually excised using the \texttt{PSRCHIVE}\footnote{\url{https://psrchive.sourceforge.net/}}~\citep{2004PASA...21..302H} software package. Each observation was then averaged in time and frequency and the pulse time of arrival was measured using \texttt{PSRCHIVE}. The analysis was carried out using the \texttt{TEMPO2}\footnote{\url{https://www.pulsarastronomy.net/pulsar/software/tempo2}} software package~\citep{2006MNRAS.369..655H} to obtain a phase-connected timing solution.

    \subsection{$\gamma$-ray motivated targets}

        \begin{table}[t]
        	\caption{\textit{Chandra}-\textit{Fermi} possible associations.}
            \begin{center}
                \begin{tabular}{c|c}
                    \hline
                    \hline
                    2CXO name & 4FGL-DR2 association \\
                        \hline
                    J173545.5--302859 & J1735.7-3026 (Ter 1) \\
                    J174235.4--282829 & J1742.5-2833 \\
                    J175039.6--302056 & J1750.4-3023 \\
                    \hline
                \end{tabular}
            \end{center}
            \label{tab:fermi_assoc}
        \end{table}
        
        A subset of the candidates\footnote{Respecting the aggressive X-ray spectral constraints of \cite{2021PhRvD.104d3007B}} identified in \cite{2024A&A...690A.330B} were also crossmatched with \textit{Fermi}'s Fourth Source Catalog \citep{2020ApJS..247...33A}. Two fall within the 95\% confidence level error circles of 4FGL-DR2 cataloged sources with no associated counterparts, and one falls within the 95\% confidence level error circle of 4FGL J1735.7-3026, associated with the globular cluster Terzan 1. We note that counterparts of \textit{Fermi} sources recorded in the Fourth Source Catalog are found using a Bayesian or a Likelihood Ratio method, solely based on spatial coincidence \citep{2010ApJS..188..405A}. All three \textit{Fermi} counterparts, listed in Table \ref{tab:fermi_assoc}, show a pulsar-like spectral energy distribution.

        \subsubsection{Observations}

        The observations of the $\gamma$-ray motivated targets were carried as part of the Transients and Pulsars with MeerKAT (TRAPUM\footnote{\url{https://www.trapum.org/}}) Large Science Project \citep{Stappers:2018Le}. The \textit{Fermi} working group, within TRAPUM, observed two of the candidates on May 20th, 2023 for 90 min each using the MeerKAT S-band receiver at central frequency 2843~MHz (S3 sub-band) and with 875 MHz of nominal observing bandwidth. The observations were designed to reach flux a density threshold of about 10 $\mu$Jy for millisecond periods: 5 ms (12 ms) if 4FGL J1742.5-2833 is at 5.2 kpc (8.5 kpc), and 1 ms for 4FGL J1750.4-3023, regardless of the source possible distance. Pulsar search-mode data were recorded with an 8-bit sampling, a sampling time of 74 $\mu$s and 2048 channels. Our observations are summarized in Table \ref{tab:obs}.

        Terzan 1, and therefore 2CXO J173545.5--302859, was observed as part of the TRAPUM globular cluster pulsar survey (e.g., \citealt{Abbate+2022,Ridolfi+2022,Padmanabh+2024}). The cluster was observed on September 21st, 2023 for 2 hours using the MeerKAT S-band receivers, at a central frequency of 2406 MHz (S1 sub-band) and with 875 MHz of nominal observing bandwidth. The data were recorded with two backends in parallel. The Pulsar Timing User Supplied Equipment (PTUSE, \citealt{Bailes+2020}) was used to record two tied-array beams, both of which had their boresights located at the nominal center of Terzan 1. The first PTUSE beam was synthesized correlating the signals of all the available antennas (56, on the day of the observation), hence it allowed the maximum sensitivity but covered a fairly small sky area, as the beam full-width-half-maximum (FWHM) was approximately 3 arcsec at 2406 MHz. The second PTUSE beam was synthesized using only 38 MeerKAT antennas, all located within an area of 1 km in radius. This allowed a larger field of view, with a beam FWHM of $\sim$ 23 arcsec, at the cost of a $\sim 30$ per cent reduction in raw sensitivity. Both PTUSE beams were recorded as search-mode PSRFITS files with full-Stokes information,  a time resolution of 18.72 $\mu$s and 8-bit digitization. The 875-MHz bandwidth was recorded as 256, 3.418-MHz-wide frequency channels, coherently de-dispersed with DM = 380.68~pc cm$^{-3}$, i.e., the average DM of all known pulsars in Terzan 1 \citep{singleton+2025}.

        In parallel with PTUSE, the Filterbanking Beamformer User Supplied Equipment (FBFUSE) computing cluster was used to correlate the signals of all the 56 antennas, to produce 18 tied-array beams. These were tiled around the globular cluster's center, with a 70 per cent overlap, using an optimal hexagonal tessellation calculated by the Mosaic\footnote{\url{https://github.com/wchenastro/Mosaic}} software \citep{Chen+2021}. The beams were in turn recorded by the Accelerated Pulsar Search User Supplied Equipment (APSUSE) computing cluster as search-mode ``filterbank'' files, retaining total-intensity only, with a time resolution of 74.89 $\mu$s, and 4096 frequency channels, with no coherent dedispersion. Shortly after the observation was over, each filterbank file of the 18 beams was incoherently de-dispersed with the same DM of 380.68 pc cm$^{-3}$ and groups of 16 adjacent frequency channels were summed together, thus producing 256-channel filterbank files, so as to significantly reduce the total data volume.

        \subsubsection{Pulsation search}
      
        The search-mode data \new{collected towards each target} were searched for periodic dispersed radio pulsations using Pulsar Miner\footnote{\url{https://github.com/alex88ridolfi/PULSAR_MINER}} \citep{2021MNRAS.504.1407R}, a Python wrapper for \texttt{PRESTO} \citep{2011ascl.soft07017R}. Our analysis includes the standard radio-frequency interference (RFI) excision, Fourier-domain acceleration search and optimization for signals with changing apparent spin periods caused by orbital motion. We account for DMs in the range 10 to 1200~cm$^{-3}$~pc.

\section{Results}
\label{sec:obs_res}

    We report on the discovery of two new radio MSPs, PSRs J1740--28 and J1740--2805, in the direction of target 2CXO J174007.6--280708. \new{Their spin periods are shorter than 30~ms, a cut-off value commonly used to disentangle MSPs from non-MSPs \citep[see e.g.][]{2023ApJ...958..191S}, but are higher than the mean or median of known the MSP population, indicating that they might not be fully recycled.} Their integrated pulse profile and frequency-phase waterfall plots are shown in Figure~\ref{fig:pulse_profiles} and their \new{preliminary} parameters are summarized in Table \ref{tab:psr}. \new{More observations are required to fully characterize these two new MSPs and understand their formation scenarios.} Both sources are further detailed below. Non-detections are also discussed and upper limits on pulsations are shown in Table~\ref{tab:ul} and Appendix \ref{app:nondet}.

    \subsection{Detections}

    \subsubsection{PSR J1740--28}
    \label{sec:det_08}

        PSR J1740--28 was first detected with 13$\sigma$ significance during a 3-hour, GBT observation at S-band (1600--2400 MHz). Subsequent observations conducted with the Murriyang telescope consistently detected the pulsar and confirmed its discovery. PSR J1740--28 has a period of 15.4~ms and a DM of $317$~cm$^{-3}$\,pc. For this line of sight and DM, the free electron density models of \cite{2017ApJ...835...29Y}, hereafter YMW16, and \cite{2002astro.ph..7156C}, hereafter NE2001, predict a distance of 5.29 and 4.58 kpc, respectively. \new{Due to the faintness of the pulsar, the time of arrivals collected during this initial timing campaign did not allow us to obtain a phase-connected timing solution. We nonetheless obtained a non-phase-connected timing solution and a rough position estimate by applying a prior on the position of 133 arcseconds, which corresponds to the radius of the GBT beam at S-band. Our analysis also revealed that PSR J1740--28 has a low-mass companion ($<0.53$~M$_\odot$) and the binary system has a period of  about 191 days and a low eccentricity of 0.0004 (see Tab. \ref{tab:psr} for more details).} 

    \subsubsection{PSR J1740--2805}
    \label{sec:det_05}

        PSR J1740--2805 was first detected in Murriyang follow-up observations of PSR J1740--28, in the upper part of the covered UWL frequency band, and later re-detected in the initial GBT data. It has a period of 7.51~ms and a DM of $330$~cm$^{-3}$\,pc. For this line of sight and DM, the YMW16 and NE2001 free electron density models predict a distance of 7.01 and 4.74 kpc, respectively. \new{We note that, according to the YMW16 distance, this MSP is very likely to be in the bulge, although we cannot completely exclude that it is in the disk. A quantitative estimate of the probability of this MSP to be in the bulge requires a theoretical work that falls beyond the scope of this observational paper.} We obtained a timing-derived position of sub-arcsecond precision of PSR J1740-2805 and found the pulsar as a continuum, circularly-polarized \new{unresolved} radio source identified as a pulsar candidate by \cite{2024ApJ...975...34F} in MeerKAT L-band imaging data, 20$^{\rm th}$ source in their Table 3. This source has reported L-band flux density of $0.86\pm0.02$ mJy and a spectral index of $-1.46\pm0.15$. \new{This result matches our VLA observations described in Appendix~\ref{app:vla}}. It was independently observed by the Survey of COmpact sources for Pulsars and Exotic objects\footnote{\url{http://www.ncra.tifr.res.in/~ymaan/scope.html}} and PSR re-detected two years after our initial discovery. PSR J1740--2805 has a low-mass companion \new{with minimal mass $0.05$~M$_\odot$} and the binary system has a \new{compact orbit with a} period of 0.48 days. \new{PSR J1740--2805 could therefore be classified as spider pulsar, more precisely a black widow, following the recent definition provided in the SpiderCat\footnote{\url{https://astro.phys.ntnu.no/SpiderCAT/}} \citep{2025ApJ...994....8K}. We note that the criteria used for classification of pulsars as spiders vary across different works, e.g. \cite{1988ApJ...335L..67E,2013IAUS..291..127R}.} Unlike PSR J1740-28, PSR J1740-2805 shows a multiple-component pulse profile, as can be seen in Figure \ref{fig:pulse_profiles}.

        PSR J1740--2805 is located 1.6 arcmin from the initial X-ray target, which indicates that they cannot be associated. No other X-ray source can be associated with the pulsar either, including the LMXB IGR J17407--2808 \citep{2023A&A...674A.100D}. The sensitivity of \textit{Chandra} at the position of the pulsar, and therefore, the upper limit on the X-ray flux of a potential X-ray counterpart is $4.62\times 10^{-15}$~erg\,cm$^{-2}$\,s$^{-1}$ in the broad band (0.5--7~keV). No \textit{Gaia} or \textit{VVV} sources are found within 2 arcsec of the pulsar position.

    \begin{table*}[t]
        \caption{Measured and derived parameters of two new MSPs.}
        \begin{center}
        
        \begin{tabular}{lcc}
        \hline
        \hline
                  & J1740$-$2805    &   J1740$-$28  \\
        \hline
        Right ascension, RAJ (J2000)        & 17:40:06.\new{7763(10)}    &   \new{17:39:58(5)}    \\
        Declination, DECJ (J2000)       & $-$28:05:33.\new{34(19)}   &   \new{$-$28:07(5)}   \\
        Spin frequency, $\nu$ (Hz)         & 133.216296979\new{09(18)}     &   \new{64.909324(3)}   \\
        Frequency derivative $\dot{\nu}$ (Hz/s) & $-2.26(\new{4})\times10^{-15}$ & \new{$-1.2(14)\times10^{-13}$} \\
        Epoch of period (MJD)       & 60438.606047113964  & \new{60418.530932677400}\\
        Dispersion measure, DM (cm$^{-3}$\,pc) &  330         &    317\\
        Time span (MJD)    &   \new{60216.34 -- 60753.80}    &  \new{60221.35 -- 60753.74}\\
        \new{1.28 GHz flux density ($\mu$Jy)} & \new{860$\pm$20} & \new{---} \\
        \new{Spectral index} & \new{--1.46$\pm$0.15} & \new{---} \\
        \new{Circular polarization fraction ($V/I$)} & \new{0.3} & \new{---} \\ 
        \hline
        \multicolumn{3}{c}{Binary parameters~\citep[ELL1 model;][]{2001MNRAS.326..274L}}\\
        \hline
        Orbital period PB (day)           & 0.48377932(\new{3}) & \new{190.60(7)} \\
        Projected semimajor axis, A1 (lt-s)          & 0.24852(\new{3})    & \new{49.09(9)}  \\
        Epoch of ascending node, TASC (MJD)         & 60439.229880(\new{8}) & \new{59896.3(3)}\\
        EPS1, $\eta$ (10$^{-3}$)   &    0.12(\new{11})    & \new{$-$0.3(13)}\\
        EPS2, $\kappa$ (10$^{-3}$) &    $-0.01$(\new{8}) & \new{0(3)}  \\
        \new{Min. and max. companion mass ($M_\odot$)} & \new{0.05, 0.12} & \new{0.20, 0.53} \\
        \hline
        \multicolumn{3}{c}{Derived parameters}\\
        \hline
        Spin period, $P$ (ms)                 & 7.5065890786383 &  \new{15.406107}\\
        Period derivative, $\dot{P}$ (s\,s$^{-1}$)  & $1.26\times10^{-19}$ & \new{$2.9\times10^{-17}$}\\
        Characteristic age, $\tau_{\rm c}$ (Gyr)     & 0.947 & 0.0085 \\
        Surface dipole magnetic field strength, $B_{\rm s}$ (G)        & $9.83\times10^{8}$  & $2.1\times10^{10}$ \\
        YMW16 distance (kpc) & 5.29 & 7.01\\
        NE2001 distance (kpc) & 4.58 & 4.74\\
        \hline
        \end{tabular}
        \tablefoot{The binary model assumes a neutron star mass of 1.4 $M_\odot$, and the minimum (maximum) companion mass corresponds to an inclination angle of $90\degree$ ($26\degree$). The flux density, spectral index and circular polarization fraction of PSR J1740-2805 are taken from \cite{2024ApJ...975...34F}. We quote errors at a 95\% confidence level.}
        \end{center}

        \label{tab:psr}
    \end{table*}

    \subsection{Non-detections}

        No pulsars were discovered towards our 8 other targeted candidates. To understand if our observations rule out their supposed nature, we provide estimated upper limits on their spin period by calculating the theoretical, maximal period detectable given their radio flux density. For radio-motivated targets, we estimated their flux density at our observing frequency from the VLA L-band imaging data in which they were initially detected. We do not provide upper limits for our $\gamma$-ray motivated targets due to the lack of identified radio counterparts. We emphasize that, as noted below, these limits depend on a number of assumptions and models, and are only given on an indicative basis. \new{For example, we cannot rule out the possibility that these sources, especially those only observed once, are pulsars in highly-accelerated and/or eclipsing binary systems. Spider pulsars, and in particular red backs, are well-known X-ray sources and their binary motion can cause eclipses.}
        
        \subsubsection{Methodology}
        \label{sec:ul_method}

        We assume that the \textit{Chandra}/VLA sources we observed have a power-law spectrum at radio wavelengths with a steep spectral index $\alpha$ as it is usually the case for MSPs.
        We extrapolate their flux densities from L-band (provided in Tab. \ref{app:J1743} and in \cite{2024A&A...690A.330B}, Tab. 1) to our GBT observing frequencies. Then, we compute periods $P$ that verify the radiometer equation, e.g., \cite{1984bens.work..234D}:
        \begin{equation}
            S(\nu) = \frac{\sigma T_{\rm tot}}{G \sqrt{n_{\rm p} T_{\rm obs} \Delta \nu}} \sqrt{\frac{w_{\rm obs}}{P - w_{\rm obs}}}
            \label{eq:radiometer}
        \end{equation}
        where $\sigma$ is the desired signal-to-noise ratio, $T_{\rm tot} = 20~\mathrm{K} + T_{\rm sky}$ is the sum of the system and the sky temperatures, $G=2$~K/Jy is the telescope gain, $n_p$=2 is the number of polarizations, $T_{\rm obs}$ the observing time in s, $\Delta \nu$ the bandwidth in MHz and $w_{\rm obs}$ is the observed width of the pulse in ms. We extrapolate the sky temperature at the source positions from the 408 MHz map of \cite{1982A&AS...47....1H}, assuming a power-law rescaling to the frequency of interest with index --2.6 \citep{1987MNRAS.225..307L}. The observed width of the pulse $w_{\rm obs}$ is computed as:
        \begin{equation}
            w_{\rm obs} = \sqrt{w^2+(\tau_{\rm scat})^2+(\tau_{\rm DM})^2+(\tau_{\rm samp})^2}
        \end{equation}
        where $w = 0.2 \, P$, assuming a 20\% pulse duty cycle, $\tau_{\rm scat}$ is the scattering time which depends on the DM at the MSP position, e.g., \cite{2004ApJ...605..759B}:
        \begin{equation}
            \begin{split}
                \log_{10}(\tau_{\rm scat}) = &-6.46 + 0.154 \log_{10}(DM)\\&+1.07 \log_{10}(DM)^2 -3.86 \log_{10}(\nu/1000 \rm MHz)
            \end{split}
            \label{eq:scat}
        \end{equation}
        and $\tau_{\rm DM} = 8.3 \times 10^6 DM \Delta\nu / (n_{\rm chan} \nu^3)$ is the DM smearing time, where $n_{\rm chan}$ is the number of channels over the bandwidth. Finally, $\tau_{\rm samp}$ is the sampling timescale.

        The values $P_\mathrm{max}$ of $P$ that verify Eq. \ref{eq:radiometer} can be interpreted as upper limits on the periodicity of undetected pulsars. We emphasize that these limits are theoretical and do not account for e.g. scintillation, eclipses, RFI and limited computing power, and assume an intrinsic pulse width, an electron density model, a distance to the source, a spectral index, and that all the radio emission comes from the pulsed emission. We also note that Eq. \ref{eq:scat} is an estimation of the scattering timescale and is subject to large uncertainties.

        \subsubsection{Upper limits}

        Using the methodology presented in Sec. \ref{sec:ul_method}, we rule out the pulsar nature for 2CXO J173946.6--282913 and J174007.6--280708, as our observations should have revealed any period larger than 0.9~ms, which is shorter than the fastest known pulsar \citep{2006Sci...311.1901H}. We rule out the young pulsar nature, but not the MSP nature, of 2CXO J174053.7--275708, 2CXO J174011.5--283221 and 2CXO J174017.3--282843, which could still have periods less than 1.52, 1.63 and 2.86~ms, respectively. Our limits on these targets are summarized in Table \ref{tab:ul}. Additionally, we can rule out the young pulsar nature of 2CXO J174306.8--293344 under certain conditions, depending on the assumptions made on its radio spectral index (not too steep) and/or its distance (closer than 8.5 kpc). More information on the spectral index and position dependency of all our upper limits is provided in the Appendix \ref{app:nondet}.

        \begin{table}[t]
            \caption{Spin period upper limits.}
            \begin{center}
                \begin{tabular}{c|c}
                    \hline
                    \hline
                    2CXO name & $P_\mathrm{max}$ \\
                     & (ms) \\
                    \hline
                    J173946.6--282913 & 0.72 \\ 
                    J174053.7--275708 & 1.52 \\ 
                    J174011.5--283221 & 1.63 \\
                    J174017.3--282843 & 2.86 \\\hline
                    J174007.6--280708 & 0.90\\
                    \hline
                \end{tabular}
            \tablefoot{Spin period upper limits ($P_\mathrm{max}$) of potential pulsars associated with \textit{Chandra} sources from current non detections. The last line refers to a pulsar associated with our original target, which is neither PSR J1740-28, nor PSR J1740-2805.}
            \end{center}
            
            \label{tab:ul}
        \end{table}

        The observations we conducted with MeerKAT should have detected any pulsation down to flux densities $\sim10~\mu$Jy. We therefore exclude the pulsar nature of 2CXO J175039.6--302056 and 2CXO J174235.4--282829, although we cannot guarantee the absence of scintillation and eclipses that could bias our observations, or that they are very faint pulsars. As the data collected towards Terzan 1 were recorded in coherent de-dispersion mode, the sensitivity of the observation to pulsars with DM values much higher or lower than the globular cluster's DM \new{could be} reduced. Hence, we conclude that the 2CXO J173545.5--302859 could either be a background or foreground pulsar, or a non-pulsar source within Terzan 1. We additionally compared the position of 2CXO J173545.5--302859 with the timing-derived positions of the known pulsars in Terzan 1 \citep{singleton+2025}, but none of the 7 pulsars with timing solutions are within 20'' of the \textit{Chandra} source. 
        
\section{Discussion}
\label{sec:discussion}

    \subsection{X-ray selection of MSP candidates}

        PSRs J1740--2805 and J1740--28 were discovered in radio observations targeting 2CXO J174007.6--280708, an X-ray source identified as a bulge MSP candidate in a previous work~\citep{2024A&A...690A.330B}. Our observations are the first observations searching for MSP associated with possible X-ray counterparts of the GCE and such observations are highly relevant to evaluate the impact of this novel candidate-identification method. The VLA Director's time observation and/or the timing-derived position ruled out the association of 2CXO J174007.6--280708 with the newly discovered pulsars, making their discovery serendipitous. We emphasize that, in \cite{2024A&A...690A.330B}, we identified more than 1400 X-ray sources as MSP candidates, while only $\sim100$ MSPs are expected to be above the X-ray sensitivity of existing \textit{Chandra} observations \citep{2021PhRvD.104d3007B}. Therefore, we expect 7\% or less of the candidates to be actual MSPs. The fraction of sources that were observed in this work represents less than one percent of all our candidates and cannot yet rule out our selection strategy. Deeper radio images of the inner Galaxy will be key in formally excluding the pulsar nature of these candidates and further refining the selection criteria by understanding discriminating X-ray features.

    \subsection{Need for deep pulsation searches}

        Existing radio surveys have been shown to be too shallow to unveil a population of Galactic bulge MSPs responsible for the GCE \citep{2016ApJ...827..143C}. Indeed, radio light coming from the Galactic center is highly dispersed \new{but most importantly so scattered that} the pulsed emission of fast pulsars is easily smoothed out in short observations. \new{One can get around dispersive smearing, to a point, by using coherent de-dispersion, but to mitigate scattering, it is necessary to observe at much higher radio frequencies, which in turn requires higher sensitivity.}
        
        The non-association of 2CXO J174007.6-280708 with either PSRs J1740--2805 or PSR J1740--28 makes the discovery of these two new pulsars serendipitous. However, it demonstrates the power of, and the need for, deep targeted radio observations. Recently, PSR J1744--2946, another MSP, has also been serendipitously discovered in deep observations targeting an X-ray source at the Galactic center \citep{2024ApJ...967L..16L}. We also note that, radio imaging with a focus on compact, polarized and/or steep-spectrum\footnote{\url{https://dlakaplan.github.io/steepspectrum.html}} sources \citep{2017MNRAS.468.2526B, 2024ApJ...967L..16L, 2024ApJ...975...34F}, has proven to be a productive and complementary approach for identifying promising bulge MSP candidates.

    \subsection{A new globular cluster?}

        The detection of two MSPs in a single radio beam, with comparable DMs, naturally raises the question of the discovery of a new globular cluster. We argue that PSRs J1740--2805 and J1740--28 are instead part of the Galactic field. Hints that PSR J1740--2805 and PSR J1740--28 belonged to a globular clusters would be, e.g. an even smaller difference in their DMs \cite[see][Figure 4]{2005ApJ...621..959F}, or negative period derivatives. Indeed, 93\% of MSPs in the ATNF pulsar catalog \citep{2005AJ....129.1993M} that have a negative period derivative (not corrected for Shklovskii effect) are in globular clusters, and 45\% of MSPs in globular clusters have negative period derivative.

\section{Conclusion}
\label{sec:conclusion}

    In this article, we report the detection of two new MSPs towards the Galactic center. PSRs J1740--2805 and J1740--28 have rotation periods of 7.5 and 15.4 ms, respectively, and DMs above 300~cm$^{-3}$~pc. Both are in binary systems with low-mass companions\new{, and PSR J1740--2805 additionally meets the criteria for black widows, a sub-type of spider pulsars}. We do not find evidence for known multi-wavelength counterparts, but find a cataloged radio continuum source at its position \citep{2024ApJ...975...34F}. Both pulsars were discovered in deep radio observations targeting, for the first time, an X-ray source identified as bulge MSP candidate \citep{2024A&A...690A.330B}. Despite the lack of MSP discoveries associated with our initial targets, our novel selection method cannot yet be ruled out and remains to be explored. The detection of PSRs J1740--2805 and J1740--28 double the number of MSPs discovered within the innermost $2\degree$ from the GC, and, given their distance estimates (see Sect. \ref{sec:obs_res}), they are both compatible with being part of the bulge population~\citep{2021PhRvD.104d3007B}. However, at these distances (around 5--6 kpc from Earth), the Galactic MSP population from the disk and the expected bulge overlap sizably. Disentangling the two with the low number of MSPs detected in the direction of the Galactic center is challenging, but the increasing number of discoveries may finally make it possible (Berteaud et al., in prep.).

\begin{acknowledgements}
     This work is supported by NASA under award number 80GSFC21M0002. JB, FC and MC acknowledge financial support from the Programme National des Hautes Energies of CNRS/INSU with INP and IN2P3, co-funded by CEA and CNES, from the ‘Agence Nationale de la Recherche’, grant number ANR-19-CE310005-01 (PI: F. Calore), and from the Centre National d’Etudes Spatiales (CNES). FC warmly thanks P.~D.~Serpico for enlightening discussion. SR is a CIFAR Fellow and is supported by the NSF Physics Frontiers Center award 2020265. Murriyang, CSIRO’s Parkes radio telescope, is part of the Australia Telescope National Facility (https://ror.org/05qajvd42) which is funded by the Australian Government for operation as a National Facility managed by CSIRO. We acknowledge the Wiradjuri people as the Traditional Owners of the Observatory site. This material is based upon work supported by the National Radio Astronomy Observatory and Green Bank Observatory which are major facilities funded by the U.S. National Science Foundation operated by Associated Universities, Inc. The MeerKAT telescope is operated by the South African Radio Astronomy Observatory (SARAO), which is a facility of the National Research Foundation, an agency of the Department of Science and Innovation. We thank staff at SARAO for their help with observations and commissioning. TRAPUM observations used the FBFUSE and APSUSE computing clusters for data acquisition, storage and analysis. These clusters were funded and installed by the Max-Planck-Institut für Radioastronomie (MPIfR) and the Max–Planck–Gesellschaft. This research has made use of data obtained from the Chandra Source Catalog provided by the Chandra X-ray Center (CXC). FA acknowledges that part of the research activities de- scribed in this paper were carried out with the contribution of the NextGenerationEU funds within the National Recov- ery and Resilience Plan (PNRR), Mission 4 – Education and Research, Component 2 – From Research to Business (M4C2), Investment Line 3.1 – Strengthening and creation of Research Infrastructures, Project IR0000034 – ‘STILES - Strengthening the Italian Leadership in ELT and SKA’.
\end{acknowledgements}

%
\bibliographystyle{aa} 
\bibliography{Biblio} 
%

\appendix

\section{2CXO J174306.8--293344}
\label{app:J1743}

In \cite{2024A&A...690A.330B}, we identified 5 promising MSP candidates detected at radio and X-ray wavelengths. In our proposed follow-up observation of these sources, we also included 2CXO J174306.8--293344, an additional MSP candidate detected at radio and X-ray wavelength with a possible infrared counterpart on the rim of the position error circle, that we interpreted as a false-positive match. The angular separation between the radio and the X-ray source is less than 0.6 arcsec and the radio source is compact in size. More information about the MSP candidate is provided in Table \ref{tab:J1743}.

\section{VLA observations}
\label{app:vla}

Successful observations were performed with the Karl G. Jansky Very Large Array (VLA) in D-array configuration (minimum baseline length of 35\,m, maximum baseline length of 1.03\,km) on November 24 and 25, 2023 under project code VLA/23B-312. These observations were conducted at L (1--2\,GHz) and S-band (2--4\,GHz) phasing the array for pulsar recording using the YUPPI backend in addition to the regular WIDAR correlation. The phasing calibrator was J1751-2524 and the target field pointing direction was R.A. 17h40m7.624s Dec. -28$^\circ$ 7' 8.750". The integration time for the target pointing was about 2 and 2.8 hours for L and S-band respectively. \new{No continuum radio source was detected at the position of our X-ray target 2CXO J174007.6--280708, indicating that no radio pulsar could be associated with it.} MSP J1740--2805\new{, for which we obtained a timing-derived position, was found} within the field of view, but 3' away from the pointing center. No pulsation was detected in the VLA data. The WIDAR correlator continuum data was manually calibrated using CASA 6.6.4-34, performing standard flagging, bandpass and complex gain calibration, as well as flux scaling based on 3C\,286. Images were obtained using the adaptive-scale pixel and multi-term multiscale deconvolver for L, S, and L\&S bands combined using natural weighting to maximize point source sensitivity. \new{In Fig.~\ref{fig:radio} the highest resolution S-band image is presented using natural weigthing of visibilities and using only a single n-term, including labeling of locations of MSP J1740--2805. The corresponding S-band (3.0\,GHz) peak flux at the location of J1740--2805 is 0.68$\pm$0.11 mJy/beam. Combination of both L and S-band data and using two n-terms (linear spectral fit) results in a 2.5\,GHz peak flux of 1.18$\pm$0.60 mJy/beam.
The location of J1740--2805 sits on-top of an extended radio source toward the Northern edge with an in-band spectral index combining L and S-band data of -1.45$\pm$0.30.} 

\section{Details on non detections}
\label{app:nondet}

\begin{figure*}[h!]
        \includegraphics[width = 0.3\textwidth]{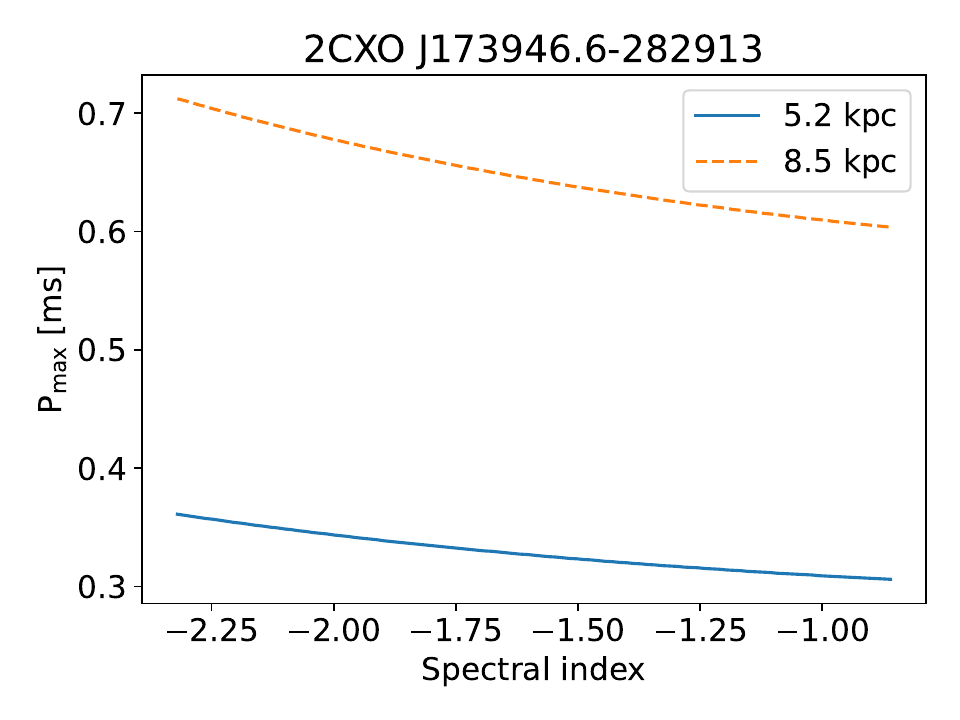}
        \includegraphics[width = 0.3\textwidth]{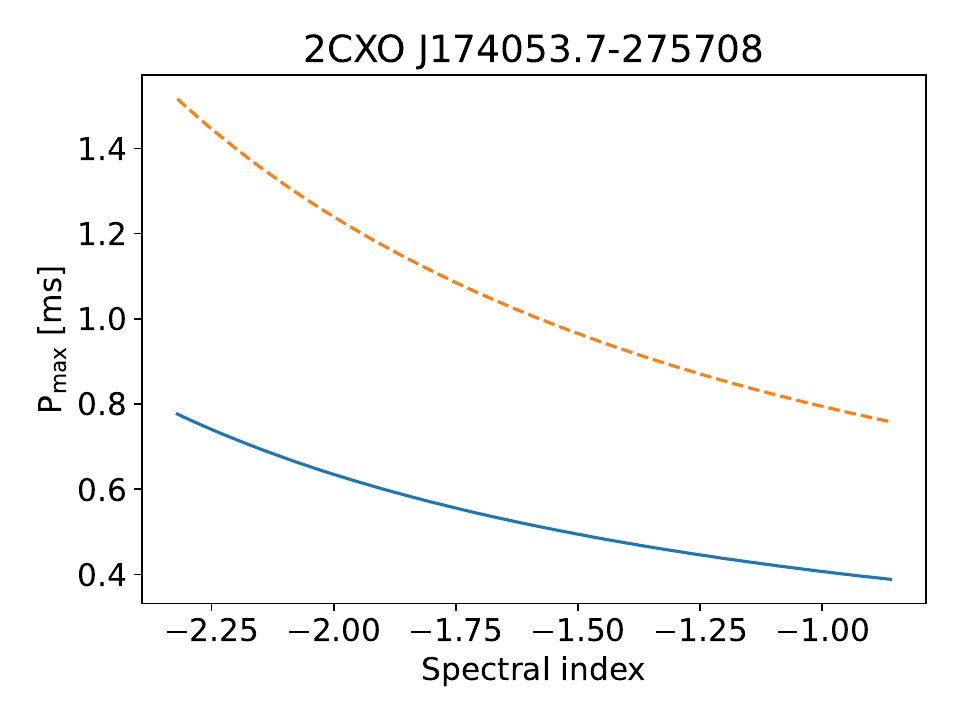}
        \includegraphics[width = 0.3\textwidth]{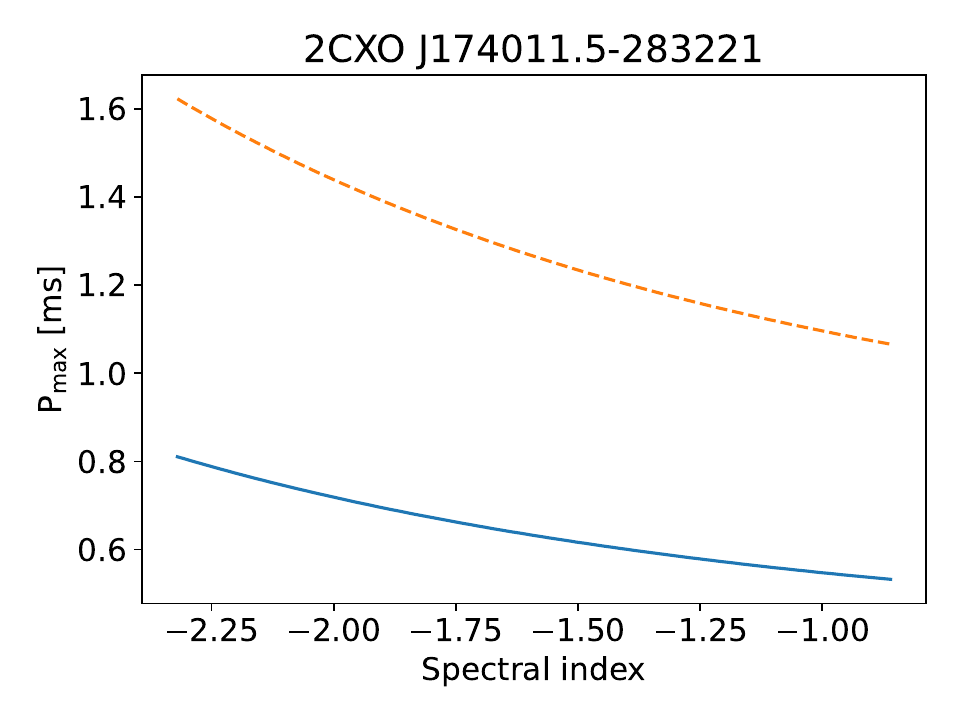}\\
        \includegraphics[width = 0.3\textwidth]{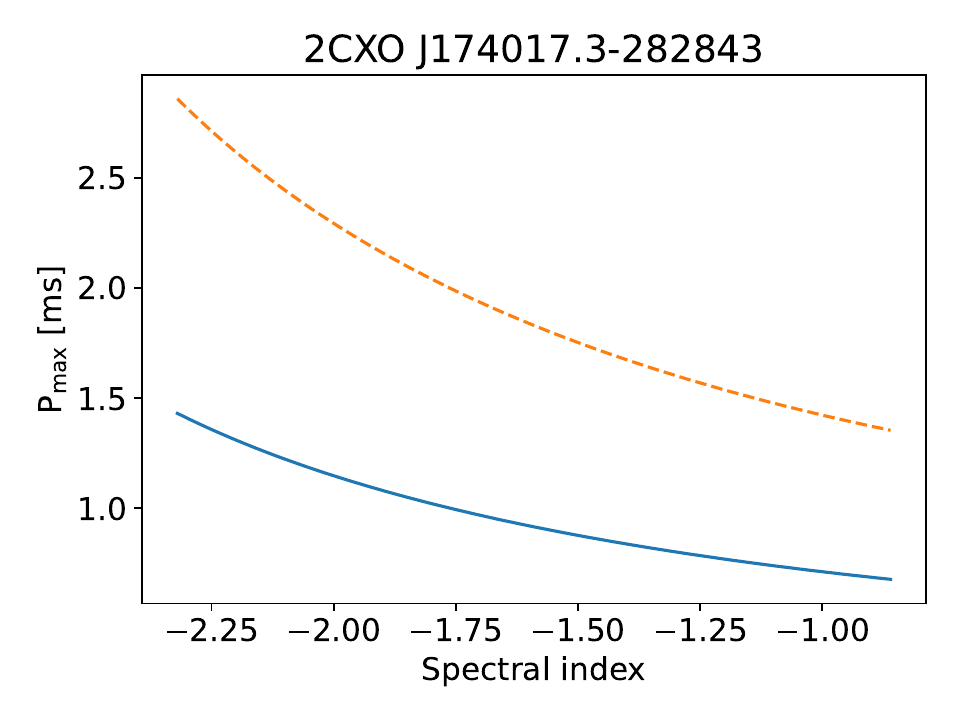}
        \includegraphics[width = 0.3\textwidth]{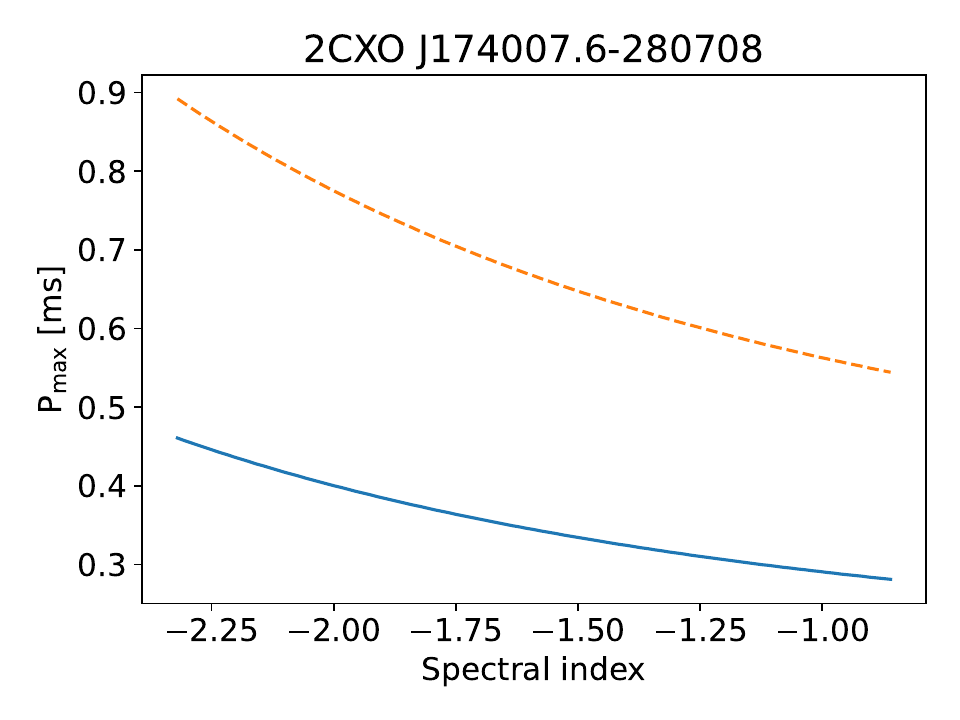}
        \includegraphics[width = 0.3\textwidth]{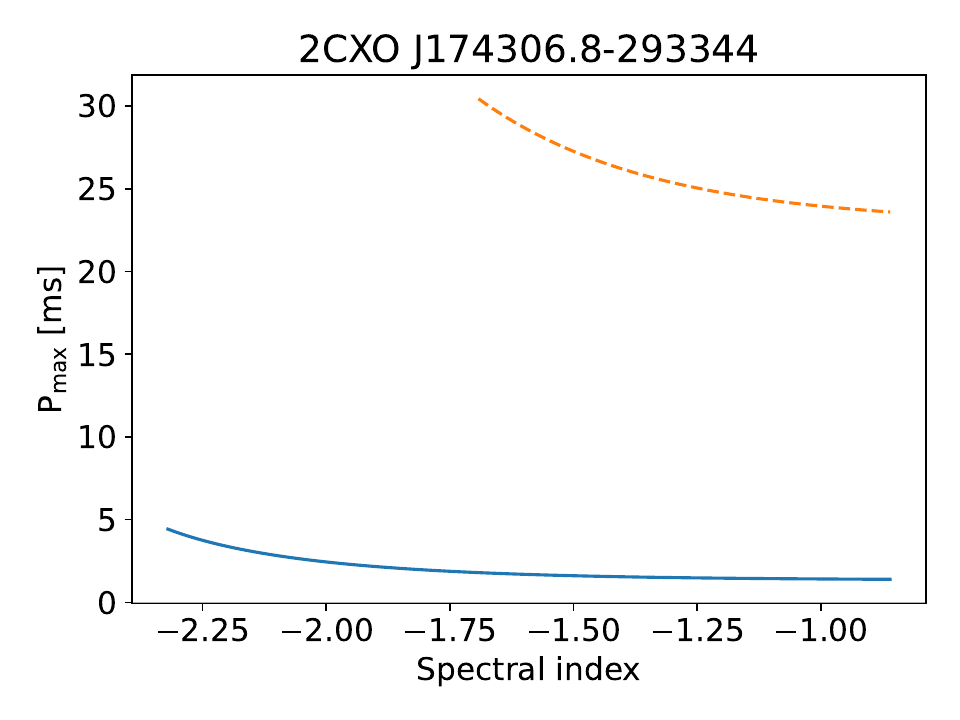}

        \caption{Maximum rotation period $P_\mathrm{max}$ for two different distances as a function of radio spectral index for our 6 MSP candidates detected by \textit{Chandra} and the VLA, and observed in this work by Murriyand and/or the GBT.}
        \label{fig:p_ul}
    \end{figure*}

\begin{figure*}[ht]
    \centering
    \includegraphics[width=0.6\textwidth]{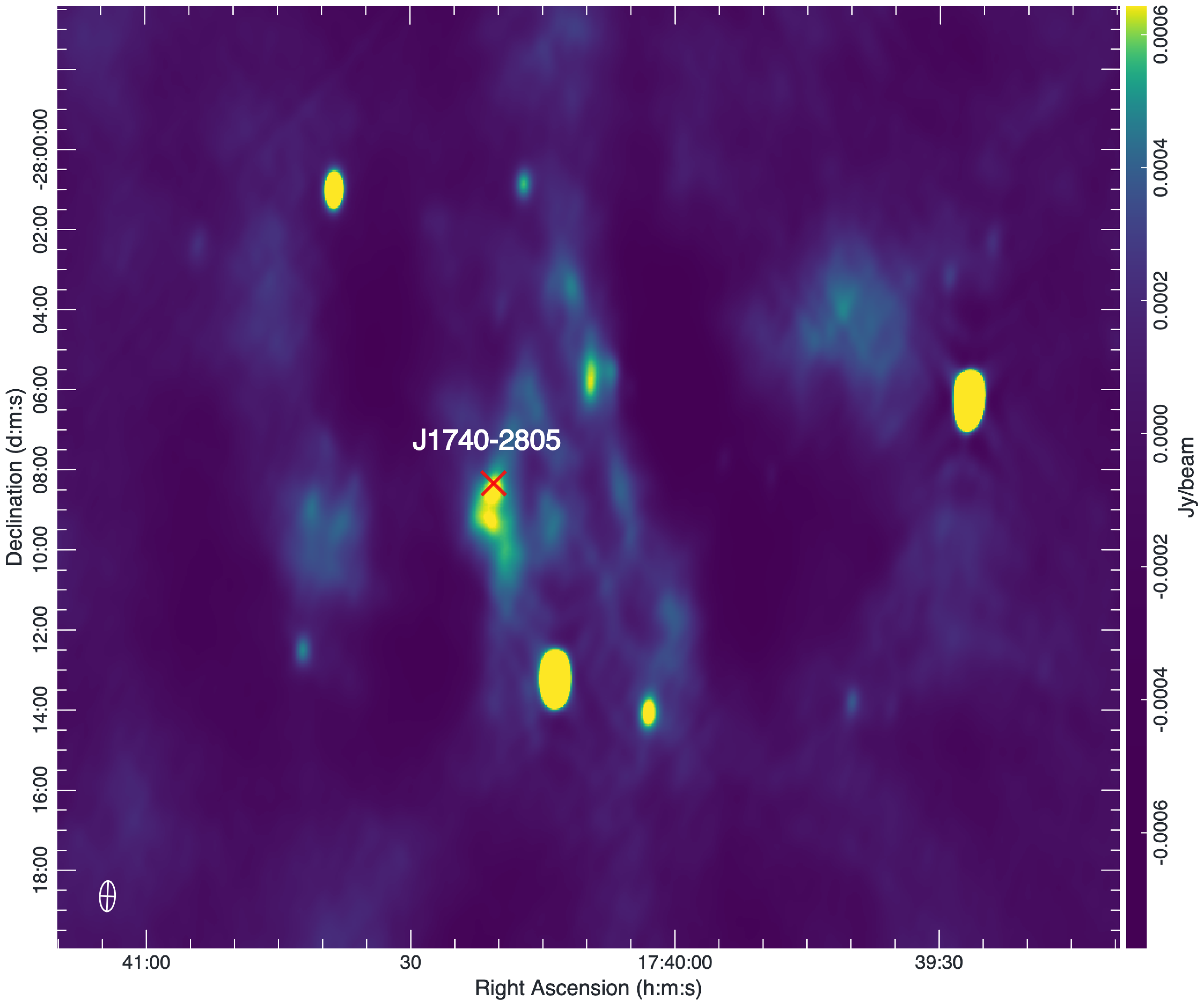}
    \caption{\new{The naturally weighted S-band radio image obtained by the VLA. The image has an rms of 100$\mu$Jy/beam. The cross marks the position of MSP J1740-2805. 
    The white ellipse in the lower left corner shows the synthesized beam shape, representing the image resolution. The color bar on the right relates the color scale to the intensity in Jy/beam.}}
    \label{fig:radio}
\end{figure*}

The detection or non-detection of a pulsar depends, for a given observation setting, on its period, flux and DM. The radio/X-ray MSP candidates observed in this work were detected in L-band imaging data but pulsations were searched at higher frequencies (S and C-band) to limit pulse broadening. Therefore, we extrapolated fluxes assuming a power-law spectrum with a steep spectral index. We estimated the DMs from the YMW16 electron density model assuming a distance of 5.2 kpc and 8.5 kpc, which are the expected minimal and mean distance of X-ray detectable bulge MSPs \citep{2021PhRvD.104d3007B}. In Figure \ref{fig:p_ul}, we show the upper limit $P_\mathrm{max}$ on the rotation period of potential pulsations as a function of the radio spectral index and the distance to the source for our 6 candidate sources, as observed with the GBT.

\begin{table*}[]
        \caption{Complement to Table 1 of \cite{2024A&A...690A.330B}.}
        \centering
        \begin{tabular}{c|ccccc}
        \hline \hline
            2CXO name & Config. & Peak flux & RA & DEC & Comment \\ 
             & & $\mu$Jy/beam & deg & deg & \\ \hline 
            J174306.8--293344 & ii & $304\pm98$ & --94.22130497 & --29.56215609 & UKIDSS source at the rim of the error circle\\
            \hline
        \end{tabular}
        \tablefoot{ From left to right, the columns give the name of the \textit{Chandra} source that matches with the VLA radio source, the configuration of the algorithm that found the source \citep[see][Section 2.3.2 for more details]{2024A&A...690A.330B}, the peak flux in L-band of the radio source and its $1\sigma$ error, its position in right ascension (RA) and declination (DEC), and comments on the multiwavelength counterparts. UKIDSS is the United Kingdom Infrared Telescope Deep Sky Survey of the Galactic plane \citep{2008MNRAS.391..136L}.}
        \label{tab:J1743}
\end{table*}

\end{document}